\documentstyle{article}
\begin{document}

\title{Theoretical confirmation of Feynman's Hypothesis on the creation
       of circular vortices in superfluid helium}

\author {E. Infeld\thanks{E-mail: einfeld@fuw.edu.pl}\ \ and A. Senatorski\\
So{\l}tan Institute for Nuclear Studies, Ho\.za 69, 00--681 Warsaw, Poland}

\maketitle

%\begin{abstract}
{\bf
The changes observed in the topology of superfluid helium vortices have
intrigued people for some time now [1]. These vortices either extend from
wall to wall, however tangled they may be in between, or else can be roughly
circular and freely move around the superfluid [1]. Some time ago, Richard
Feynman postulated that two oppositely polarized line vortices could, if they
cross at two points, reconnect so as to create a circular vortex that snaps
off and subsequently lives a life of its own [2]. This is often simply
postulated in numerical experiments, e.~g. [1], [3]. That an opposite line
vortex pair solution of the Nonlinear Schr\"odinger equation (NLS) for HeII is
unstable has been demonstrated theoretically [4]. Reconnection at a point has
been obtained numerically [5]. What remained to prove was that known,
stationary, double vortex line solutions can thus reproduce one of the one
parameter solitonic family of solutions for circular vortices as found
for NLS [6]. In other words, can this reconnection really lead to a full
confirmation of Feynman's hypothesis? Our first step was a similar changeover
calculation for a limiting case, unfortunately such that the vortex
configuration was degenerate [7], [8]. However, surprisingly complete dynamic
changeover from cylindrical to spherical symmetry of the soliton was observed
in [7]. This augured well for the present effort.}\\[3ex]
%\end{abstract}

An imperfect Bose condensate, such as HeII, can (also imperfectly) be described
by a single particle wavefunction $\psi({\bf x},t)$ of $N$ bosons of mass $m$
that obeys the
Nonlinear Schr\"odinger equation (NLS), according to Gross, Pitayevski and
Ginzburg:
\begin{equation}\label{nls}
{\rm i} \hbar \frac{\partial \psi}{\partial t} = - \frac{\hbar^2}{2m}
\nabla^2 \psi + W_0 \psi |\psi|^2 \, .
\end{equation}
Here $W_0$ characterizes the potential between bosons. Opposite vortex pair
solutions, as well as those describing circular vortices are known [6].
Each solution has a unique velocity perpendicular to the vortex plane.
However, to answer the crucial question of whether dynamics can lead from
the former to the latter kind, we must resort to numerics. All theory tells us
is that the double line vortex configuration is unstable [4].

Before giving the results of our simulations, we wish to point out that
a preliminary idea of the problem can be gained from the linear equation,
$W_0 = 0$. At a vortex, $\psi = 0$, so the cubic term not contributing locally
might not be too surprising. However, the extent of global similarities with
solutions to equation (1) may be more so. The preliminary results will
help us appreciate just what the role of the nonlinear, $W_0 > 0$, term is in
the act of reconnection.

One can easily check by substitution that equation (1), $W_0 = 0$, is solved by
\begin{equation}\label{psi}
\psi = \mbox{const}\left[ a^2 - x^2 + {\rm i} b \left( z(t) + \frac
{\hbar t}{mb} \right) \right] {\rm e}^{{\rm i} k_z z - {\rm i} \hbar k_z^2
t/(2m)} \, , \quad
z(t) = z - \frac{\hbar k_z}{m} t, \quad b > 0 \, .\nonumber
\end{equation}

Vortices are situated where both Re$\, \psi$ and Im$\, \psi$ are zero. They
constitute two oppositely polarized lines along $y$ at $x = \pm a$ and move
together at velocity $U_z = \hbar (k_z - b^{-1})/m$.
There is no correlation between the seperation, $2a$, and the
uniform velocity $U_z$, which can in fact have either sense.

A second solution to equation (1), $W_0 = 0$, is given by [9]:
\begin{equation}\label{psi2}   
\psi = \mbox{const} \left[ R^2 - x^2 - y^2 + {\rm i} d \left( z(t) +
\frac{2 \hbar}{md} t \right) \right]
{\rm e}^{{\rm i} k_z z - {\rm i} \hbar k_z^2
t/(2m)} \, .
\end{equation}
A circular vortex at $x^2 + y^2 = R^2$ is now moving up $z$ at velocity
$U_z = \hbar (k_z - 2 d^{-1})/m$. Again, there is no connection between $R$
and the uniform velocity, or even with its sense.

Jones and Roberts found both a class of stationary, double line vortex
solutions to (1), $W_0 > 0$, as well as circular ones. Correlations between
$a,\ R$ and corresponding $U_z$ were given in tables. Otherwise, the
similarities between their solutions with the above are at first surprising,
especially if we choose the velocities in (2) and (3) such as to mimic those
of Jones and Roberts. The role of the nonlinear term would then ostensibly be
limited to ensuring that $|\psi|$ tend to a uniform value in the far field.
However, there is a less obvious difference. Even if we perturbed (2) such
that two vortices touched at a point, say by adding $a \cos (k_y y)$ to $x$
initially, a circular vortex would not be produced at any $t > 0$.

Further calculations will be compared with the solutions of Jones \& Roberts.
Therefore we cast equation (1) in dimensionless form such that we can use
their tables (here $E$ is the average energy level per unit mass of a boson):
\begin{equation}\label{trans}
\psi \to {\rm e}^{-{\rm i}mEt/\hbar} \psi \, , \quad {\bf x} \to
\frac{\hbar}{\sqrt{2E} m} {\bf x} \, , \quad t \to \frac{\hbar}{2mE} t
\, ,
\end{equation}
so finally $\psi \to \sqrt{\frac{mE}{W_0}} \psi$.
(Linear models will match the temporal dependence if $k_z = \sqrt{2E}
m/\hbar$.)
Now
\begin{equation}\label{nlsdl}
2 {\rm i} \frac{\partial \psi}{\partial t} = - \nabla^2 \psi -
\psi (1 - |\psi|^2) \, .
\end{equation}
If we write $\psi = \rho^{1/2} {\rm e}^{{\rm i}S}$, then $\rho$ and
${\bf v} = \nabla S$ have a fluid interpretation. The variables $\rho$
and ${\bf v}$ satisfy the usual continuity equation, but due to the nonlinear
term, the Newtonian equation has a rather strange pressure tensor [6]. This
may explain the possibility of reconnection. Importantly, if we encircle
a $\psi = 0$ line once, $S$ must increase by $\pm 2 n \pi$ so that $\psi$
is single valued. This was the case for (2) and (3), where $n = 1$ (unless
$a = 0$ in (2), in which case $n = 0$).
At infinity, $|\psi| \to 1$ and this must be included in our initial
conditions describing the pair of line vortices.

As initial condition, we took a two vortex configuration in which the
seperation and velocity were lifted from Jones and Roberts, Table 2. Thus
\begin{equation}\label{incon}
\psi (t = 0) = \frac{r_1 r_2}{\sqrt{r_1^2 + b^2} \sqrt{r_2^2 + b^2}}
\, {\rm e}^{{\rm i}(\theta_1 + \theta_2)} \, ,
\end{equation}
where                                                                       
\[
r_1 = (1 - 2 U^2)(x + a)^2 + z^2 \, , \quad
r_2 = (1 - 2 U^2)(x - a)^2 + z^2 \, , \quad
\tan \theta_{1,2} = \frac{z}{\sqrt{1 - 2 U^2}(a \pm x)} \, .
\]
In spite of the scaling of $x$, following from the asymptotics of equation (5),
[6], $\theta_i$  increase or decrease by $2 \pi$  when a vortex is encircled
once. Note that $|\psi| \to 1$ in the far field. The constant $b$ was chosen
such that the subsequent velocity along $z$ in the simulation would agree with
that in the formula (we know from Fetter's solution [10] that $b \to 2$ as
$a \to \infty$ and $U \to 0$). Initially we took $U = 0.3$
and  $a = 1.75$ from Table 2 of [6], assuming periodic boundary conditions.
Next our initial condition was perturbed along cyclic $y$ and the dynamic
development was followed from equation (4), Fig 1. The circular vortex of
Fig. 1c was obtained. Its radius and velocity agree with those predicted by
Jones and Roberts. The circular vortex moved forward with uniform velocity and
negligible change of shape, thus confirming that it is indeed a Jones \&
Roberts solution.
We repeated the calculation for different initial conditions, always obtaining
viable circular vortex solutions, see Fig. 2. Thus, Feynman's hypothesis is
confirmed, completing the tentative steps of [5] and [7]. Of course, this
confirmation is only as conclusive as is the NLS model for a Bose gas.
With the above reservation, the experimentally found abundance of circular
vortitices in superfluid HeII is now explained theoretically. The proximity at
two points of two opposite line vortices in so tangled a web is quite
commonplace [1].
The generation of vortex rings due to the helical instability of a vortex
line is also of primary interest in superconductivity theory [11], [12].
Perhaps our experience could be useful there, though unfortunately the
equations are more complicated (in the Ginzburg--Landau model, the vector
potential ${\bf A}$ appears in an extension of (1). An additional vector
equation relates ${\bf A}$ and $\psi$).\\[2ex]

The authors would like to thank Drs T. Lenkowska Czerwinska and A. A.
Skorupski for help in preparing this text.%\\[3ex]

\noindent
FIGURE CAPTIONS\\[3ex]

\noindent
Figure 1. Three stages in a Feynman transformation of two perturbed line
vortices into a circular vortex in HeII, as follows from the NLS equation.
Densities on the axes are zero.\\[2ex]

\noindent
Figure 2. Our numerically obtained circular vortices (circles) as compared
to those of Jones and Roberts (continuous line) in $R,\ U$ space.

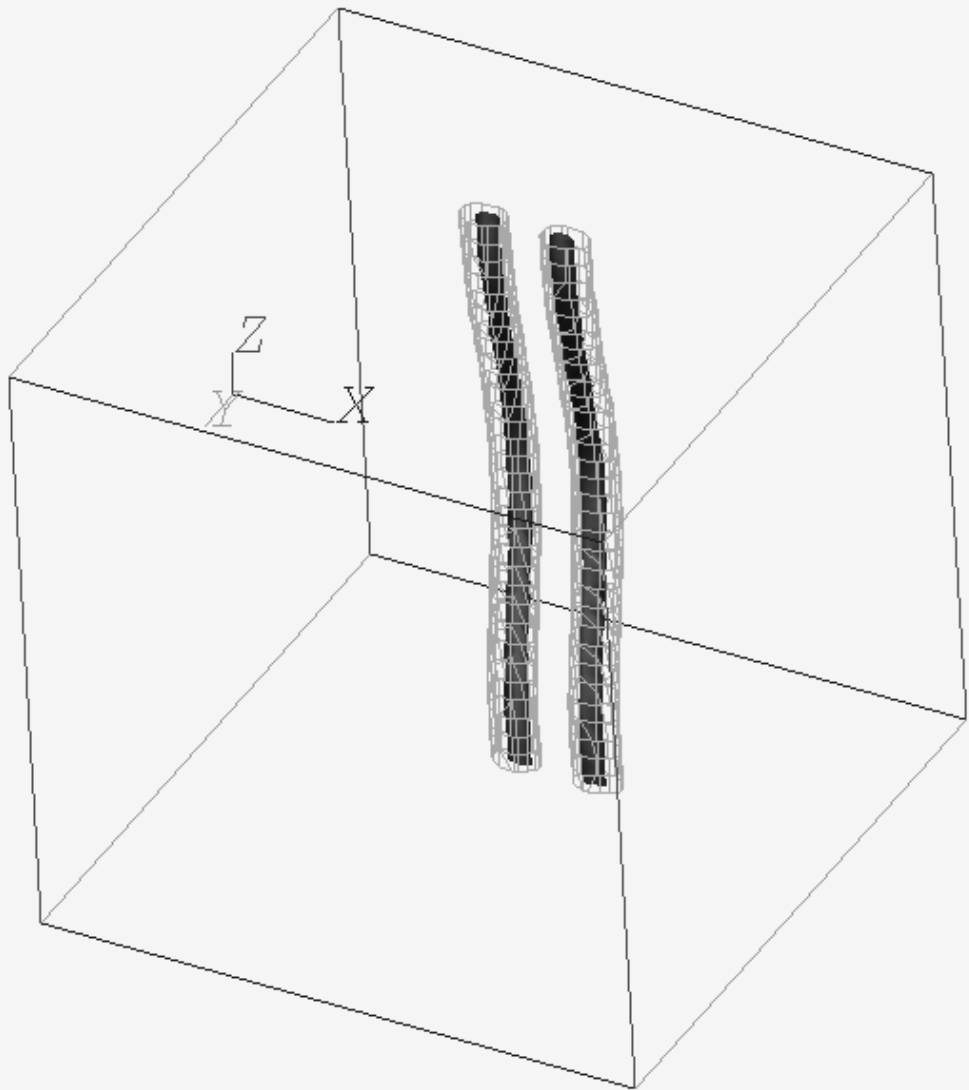
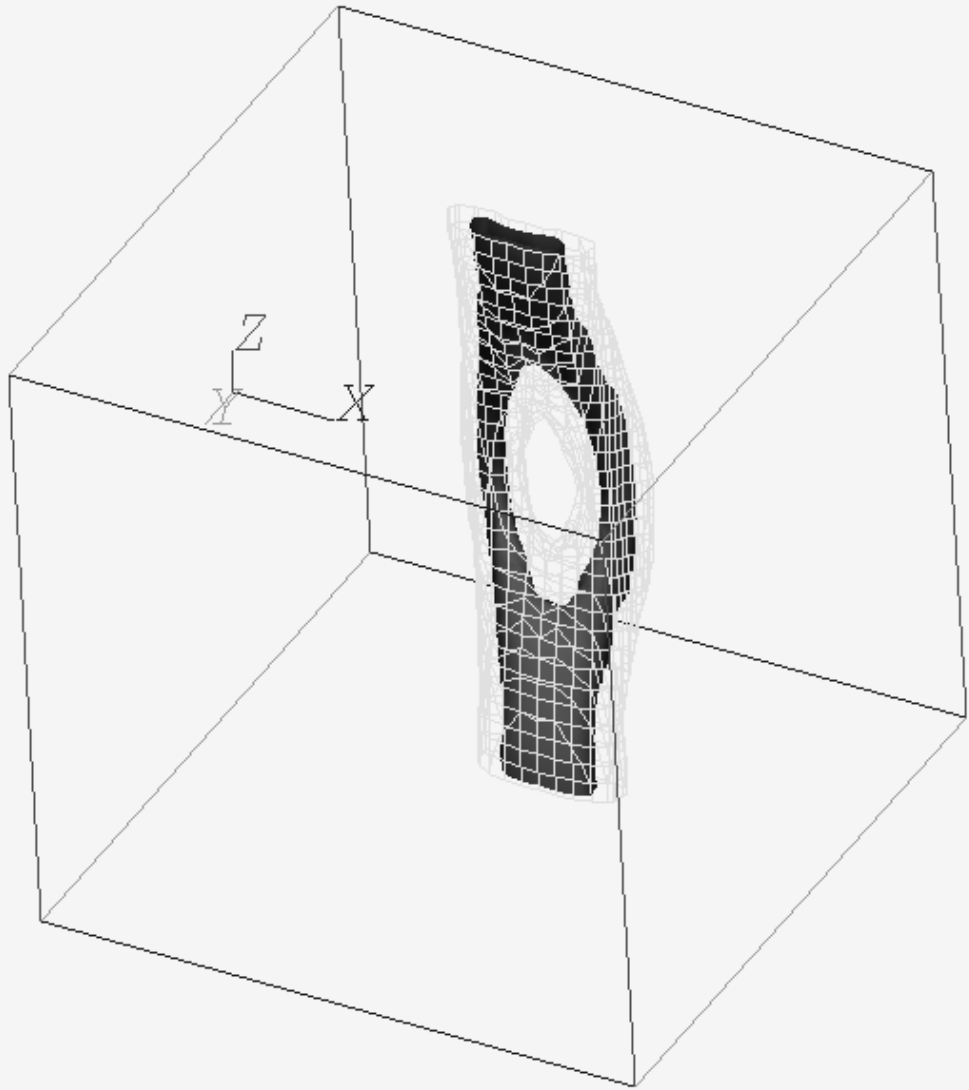
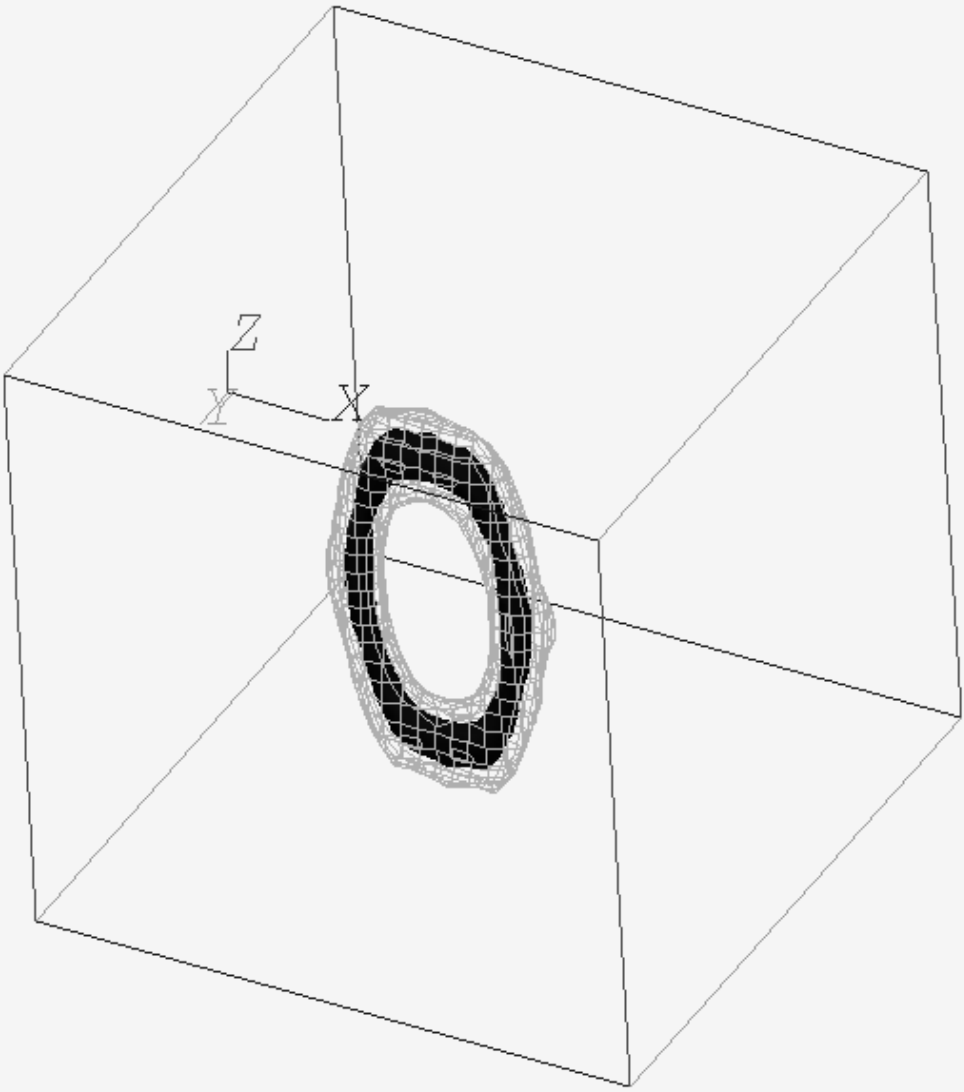
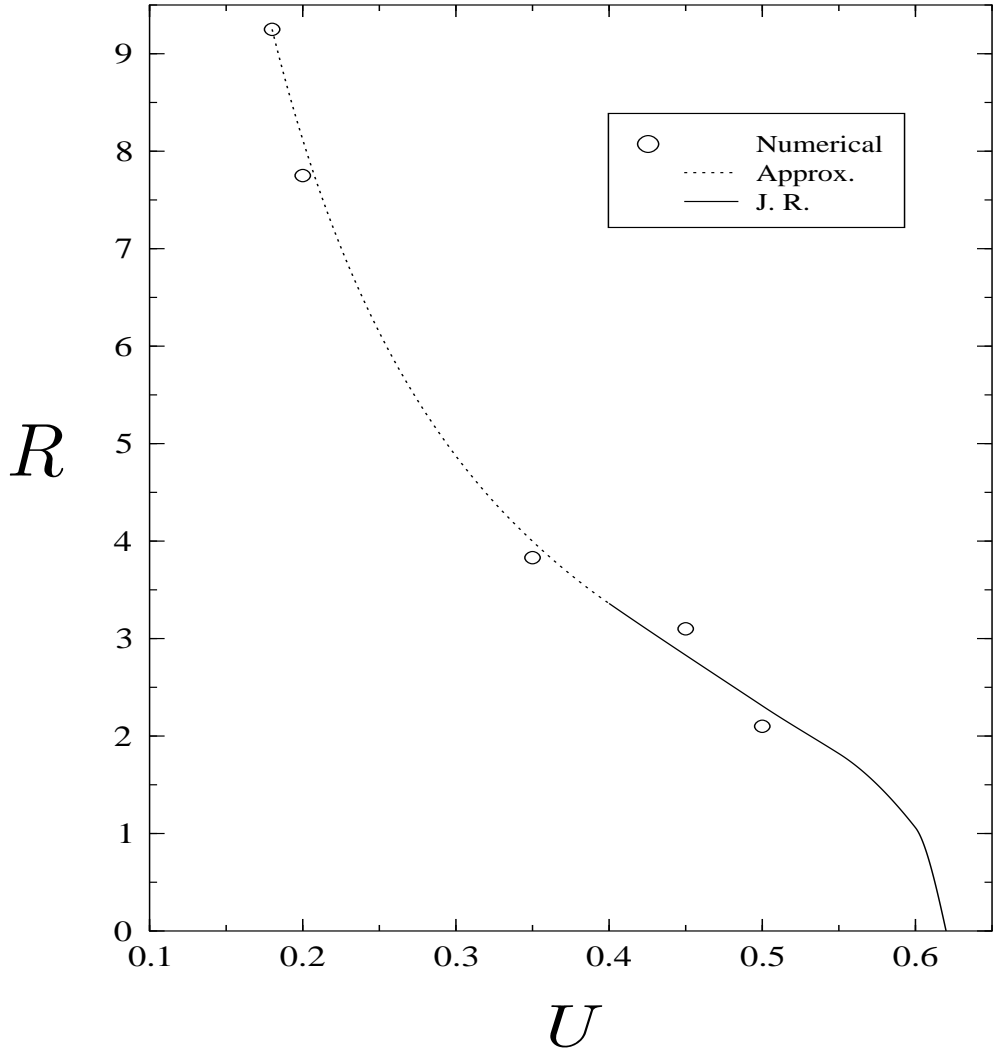
\end{document}